\documentclass[
  12pt,
  aps,
  floatfix,
  letterpaper,
  prx,
  longbibliography,
  singlecolumn,
  reprint,
  superscriptaddress
]{revtex4-2}

\AtBeginDocument{
  \renewcommand{\selectlanguage}[1]{}
}

\usepackage{newtxtext,newtxmath}

\usepackage[colorlinks=true, linkcolor=blue]{hyperref}
\usepackage{graphicx}
\usepackage{amsfonts}
\usepackage{amsmath}
\usepackage{braket}
\usepackage{gensymb}
\usepackage{dsfont}

\usepackage[capitalize]{cleveref}

\usepackage[utf8]{inputenc}

\date{\today}

\usepackage{subfiles}

\begin{document}

\begin{abstract}

Spatiotemporal fluctuations across multiple scales govern and characterize the emergent properties, phase boundaries, and low-energy excitations of strongly correlated matter. However, capturing these dynamics has remained a major experimental challenge, as
conventional probes typically offer either high spatial or temporal resolution, but not both simultaneously. Here, we leverage high-fidelity wide-field imaging of dense diamond nitrogen vacancy center ensembles to measure the momentum and frequency power spectral density of magnetic fluctuations. To access spatial wavevectors below the diffraction limit, we tune the sensing volume continuously through optical depletion. This approach enables study of equilibrium and driven fluctuations across three orders of magnitude in spatial scale and tunable frequency bands, providing a direct means to map low-energy, long-wavelength fluctuations in correlated systems.

\end{abstract}

\title{Momentum-resolved magnetic noise spectroscopy using ensembles of diamond quantum sensors}
\author{Zeeshawn Kazi}
\thanks{These authors contributed equally to this work.}
\affiliation{Princeton University, Department of Electrical and Computer Engineering, Princeton, NJ 08544, USA}
\author{Kai-Hung Cheng}
\thanks{These authors contributed equally to this work.}
\affiliation{Princeton University, Department of Electrical and Computer Engineering, Princeton, NJ 08544, USA}
\author{Jared Rovny}
\affiliation{Princeton University, Department of Electrical and Computer Engineering, Princeton, NJ 08544, USA}
\author{Nathalie P.\ de Leon}
\thanks{Corresponding author. Email: npdeleon@princeton.edu}
\affiliation{Princeton University, Department of Electrical and Computer Engineering, Princeton, NJ 08544, USA}

\maketitle

Dynamics in materials can be characterized by the dynamical structure
factor~\cite{van_hove_correlations_1954}, which quantifies correlations among degrees of freedom in a given system. Fluctuations of spin, magnetic, or charge degrees of freedom produce magnetic noise, and directly measuring this magnetic noise offers a unique window into the structure factor~\cite{casola_probing_2018, rovny_nanoscale_2024}. Nitrogen vacancy (NV) centers in diamond are exceptionally capable for magnetic noise spectroscopy because of their high-sensitivity nanoscale operation over many frequency bands and temperatures, and their solid-state host material that enables sensor-sample integration in a variety of geometries~\cite{casola_probing_2018, rovny_nanoscale_2024, pakpourtabrizi_highquality_2026, riedel_scalable_2026, guo_direct-bonded_2024}. NV center noise spectroscopy has been applied to probe phase transitions and magnon dynamics in magnetic materials~\cite{ziffer_quantum_2024, xue_magnon_2026, li_critical_2025}, Johnson noise in normal metals~\cite{kolkowitz_probing_2015} and quasiparticle and vortex dynamics in superconductors~\cite{liu_quantum_2025, li_nanoscale_2026, monge_spin_2023, jayaram_probing_2025}. Moreover, simultaneous phase measurements of two NV centers have been shown to enable measurement of two-point magnetic noise correlators~\cite{rovny_nanoscale_2022}, which could provide insight into the microscopic nature of fluctuations~\cite{zhang_nanoscale_2026, zhang_detecting_2026, hosseinabadi_theory_2026}. These correlation measurements have been extended to length scales below the optical diffraction limit~\cite{rovny_multi-qubit_2025} and frequencies from DC to GHz~\cite{le_wideband_2025}, and even further to sensing with maximally entangled states~\cite{rovny_multi-qubit_2025}, which provide a large quantum advantage for sensing
nanoscale correlators.

However, because these noise sensing measurements are performed with fixed separations, they sample the momentum response sparsely and thus do not have the capability to resolve the full momentum response of magnetic fluctuations, which could provide information about many phenomena such as topological phase transitions~\cite{curtis_probing_2024, potts_spin-qubit_2025} and the rich dynamics in superconductors~\cite{cheng_qubit_2026, dolgirev_characterizing_2022, chatterjee_single-spin_2022}. One proposal is to use a retractable tip with an NV center as a momentum-resolved magnetic noise probe~\cite{cheng_qubit_2026, agarwal_magnetic_2017}. However, momentum-resolved noise spectra have not been constructed from scanning tip measurements because moving the tip away from the material also reduces coupling to the sample~\cite{monge_spin_2023, jayaram_probing_2025}, and moving the tip close to the sample may incur additional practical challenges~\cite{xu_minimizing_2025}. Furthermore, this modality assumes that the structure factor of the noise is isotropic. Experiments scaled up to enable multiplexed correlation measurements of several shallow NV centers~\cite{cheng_massively_2025, cambria_scalable_2025} can measure many two-point correlators simultaneously, but suffer from sparse and stochastic spatial sampling that precludes the full construction of the momentum-resolved magnetic noise power spectral density.

\begin{figure*}[h!]
	\centering
	\includegraphics[width=180mm]{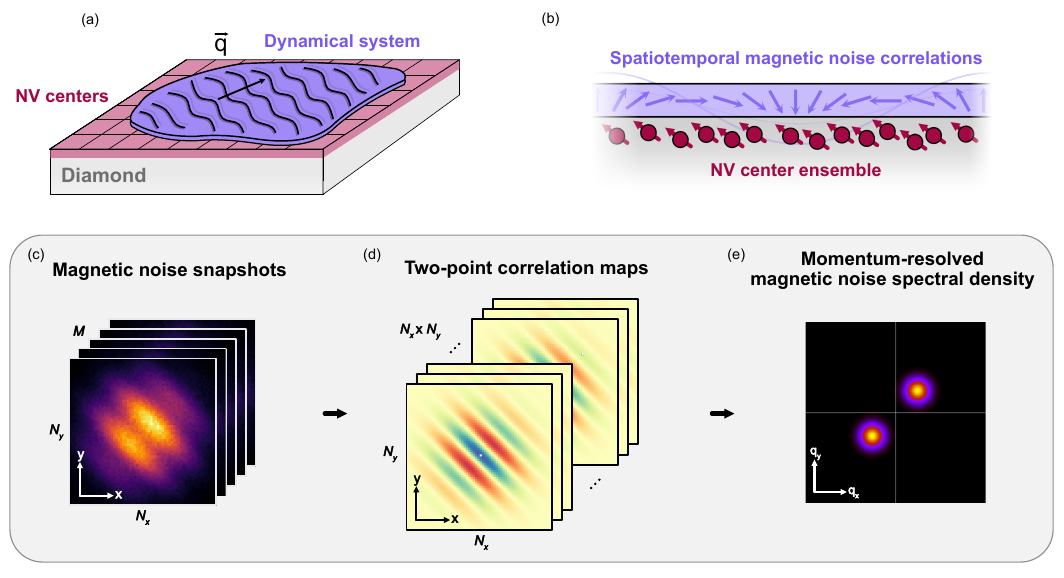}
	\caption{\textbf{Momentum-resolved magnetic noise spectral density.}
		(a) Schematic of a correlated, dynamic material system directly on top of a diamond hosting NV centers.
		(b) Zoomed in schematic showing correlated dynamics of the system sensed by a shallow NV center ensemble.
		(c) Fluorescence images record the magnetic noise measured by NV centers across the imaging field of view. Each pixel records the signals sensed by hundreds of NV centers ($N_x$ by $N_y$ pixels), and $M$ shots are taken.
		(d) High-fidelity imaging of NV center signals enables the construction of $N_x\times N_y$ maps of two-point magnetic noise correlators. Each map represents the correlation at a given pixel with respect to a reference pixel (marked in white).
		(e) The Fourier transform of the magnetic correlation function obtained from all possible correlation maps enables construction of the magnetic noise spectral density at finite spatial momenta $\mathbf{q}$.}
	\label{fig:overview}
\end{figure*}

In this work, we demonstrate methods to measure the momentum spectrum of magnetic noise. Our platform utilizes a high-density NV center ensemble in close proximity to a system~[\cref{fig:overview}(a)] that produces magnetic noise correlations over space and time~[\cref{fig:overview}(b)]. We use a noise sensing sequence to map the magnetic field at a well-defined frequency to fluorescence intensity and image using a high-speed, low-noise camera. From the stack of fluorescence images obtained from repeated measurements~[\cref{fig:overview}(c)], we compute two-point correlations among all pixels in the field of view. We illustrate this functionality by plotting maps that show the magnetic field
correlations~[\cref{fig:overview}(d)] at each position with respect to a target pixel. The measurement of all possible correlators over all separations enables us to compute the Fourier transform of the correlation function, which shows features at finite momenta corresponding to the spatial structure of the magnetic noise signal~[\cref{fig:overview}(e)]. Combining momentum resolution with methods to tune the frequency filter function~\cite{degen_quantum_2017} provides the ability to measure the full magnetic noise spectral density $S(\mathbf{q},\omega)$ in the frequency and momentum bands accessible
with NV center ensembles.

\section{Imaging two-point correlators}
The key functionality necessary to measure the spatial structure of magnetic noise is high-fidelity, wide-field imaging of two-point magnetic field correlators. To this end, we demonstrate wide-field spin-to-charge conversion (SCC) readout of NV center ensembles~\cite{shields_efficient_2015, jayakumar_spin_2018, noauthor_materials_nodate}. We fabricate a diamond with a high density of NV centers approximately 10$\pm$3~nm from the surface with $>4\%$ conversion efficiency of nitrogen ions to NV centers~\cite{healey_production_2026}, and we image the NV centers using a low-noise, high-speed camera. Each diffraction-limited pixel (250~nm) records the fluorescence from a sub-ensemble of NV centers~[\cref{fig:wfsccandcorrs}(a)]. Illumination with a high-intensity red laser induces spin-dependent ionization, and we read out the resulting charge state using a low-power orange laser. We quantify the readout fidelity by measuring the photon number distributions in the $|0\rangle$ and $|1\rangle$ spin states, from which we compute the spin state readout noise $\sigma_R$~\cite{taylor_high-sensitivity_2008, shields_efficient_2015}. The time required to sense a given correlated field strength scales as $\sigma_R^4$~\cite{rovny_nanoscale_2022}. By treating each pixel as an individual sensor, we obtain a regular array of low readout noise sensors~[\cref{fig:wfsccandcorrs}(b)]. The minimum sensor size is given by the diffraction limit, and the lateral size of the sensing array is limited by the available power for spin-dependent ionization~\cite{noauthor_materials_nodate}. We additionally measure the scaling of readout noise as a function of number of NV centers in the sub-ensemble and compare our wide-field SCC readout to conventional green readout in \cref{fig:wfsccandcorrs}(c). We find that SCC strongly outperforms green for wide-field correlation sensing with diffraction-limited spatial resolution, but the similar readout noise and faster readout time enables comparable sensitivity when using around 10$^5$ NV centers~\cite{noauthor_materials_nodate}. Thus green readout can be leveraged for sensing correlated noise over long length scales, where the significantly lower laser power requirements for green readout also enable measurements of much larger fields of view.

Equipped with high-fidelity, wide-field readout, we turn to measuring two point correlators among all sub-ensembles simultaneously. To illustrate this, we first perform a driven correlation experiment~\cite{cheng_massively_2025, cambria_scalable_2025}, in which we alternate flipping all sensors between the $|0\rangle$ and $|1\rangle$ spin states, optically read out the spin states using wide-field SCC, and then compute correlation coefficients between all pixels. We plot the resulting correlations in a map~[\cref{fig:wfsccandcorrs}(d)], where the image value represents the Pearson correlation coefficient of each pixel with respect to the pixel marked in white. We use a separate laser to flip the spin state of a sub-set of pixels (dashed circle), causing them to flip out of phase with the rest of the field of view, meaning the correlations between sensors prepared in the same state as the white pixel (blue) are positive, and between sensors prepared in the opposite states (red) are negative.

We then turn to wide-field sensing of correlated magnetic noise~\cite{rovny_nanoscale_2022}. We apply global noise from a phase-random RF test tone and use XY8 dynamical decoupling to sense the correlated noise across the imaging field of view~\cite{noauthor_materials_nodate}.  The high-fidelity readout makes correlations visible in scatter plots of the mean-subtracted photon counts measured at many pixels simultaneously~\cite{zheng_differential_2022}: correlations and anti-correlations caused by applied noise manifest as ellipsoidal photon distributions [Fig.~\ref{fig:wfsccandcorrs}(e)], where the correlation is quantitatively given by the slope of the major axis and the variances of the individual photon distributions~\cite{noauthor_materials_nodate}. When the sensing sequence is off resonance with the applied noise, no correlation is observed and the slope vanishes.

\section{Magnetic noise in momentum space}
The massively-multiplexed measurement of two-point magnetic field correlators enables construction of the magnetic noise power spectral density with momentum resolution, which is directly related to the dynamic structure factor~\cite{noauthor_materials_nodate}. By the Wiener-Khinchin theorem, the momentum-resolved power spectral density is defined as the Fourier transform of
the two-point correlation function:

\begin{equation}
	S(\mathbf{q},\omega_0)=\frac{N\pi}{\omega_0}\int_{-\infty}^{\infty}\,d^2\boldsymbol{\ell}\,e^{-i\mathbf{q}\cdot\boldsymbol{\ell}}\,\langle B_{\omega_0}(\mathbf{x}_i)B_{\omega_0}(\mathbf{x}_j)\rangle
	\label{eq:sqw}
\end{equation}

where $\langle B_{\omega_0}(\mathbf{x}_i)B_{\omega_0}(\mathbf{x}_j)\rangle$ is the magnetic field correlator measured at frequency $\omega_0$ with $N$ dynamical decoupling pulses, and $\boldsymbol{\ell}=\mathbf{x}_i-\mathbf{x}_j$ represents all possible separations~\cite{noauthor_materials_nodate}. We first demonstrate the construction of $S(\mathbf{q},\omega_0)$ for globally correlated noise. The real-space correlated noise data can be visualized with a set of two-point magnetic correlator maps~[\cref{fig:sqw}(a)]~\cite{cheng_massively_2025}. We then apply \cref{eq:sqw} to construct $S(\mathbf{q},\omega_0)$~[\cref{fig:sqw}(b)]. Because the noise is globally correlated, its magnetic noise power spectral density appears as a $\mathbf{q}=0$ feature, broadened into a Gaussian whose momentum-space width is inversely proportional to the spatial extent of the imaging field of view. The radially-averaged profiles in \cref{fig:sqw}(c) show the effect of changing noise amplitude on the magnetic noise power spectral density, and the integrated magnetic noise power (inset) shows the expected functional scaling with increasing applied noise~\cite{rovny_nanoscale_2022}.

Crucially, our technique extends beyond isotropic and long-wavelength profiles, enabling us to learn the spatial structure of magnetic noise signals with anisotropic, periodic fluctuations. To demonstrate this, we fabricate an array of parallel wires, each 1~µm wide and separated by 1~µm, and place a diamond microchip hosting a high-density ensemble of NV centers~\cite{riedel_scalable_2026} on top of the wire array~[\cref{fig:sqw}(d)]. The geometry of the wire array introduces a spatial modulation of the magnetic field amplitude produced by the applied current with periodicity given by the wire separation. We measure the momentum-resolved magnetic noise spectral density with and without applied current, and plot their difference in \cref{fig:sqw}(e). We observe two sharp features in the map, and a line cut through the features~[\cref{fig:sqw}(f)] reveals peaks at $q_{||}\approx2\pi\times0.5$~µm$^{-1}$, corresponding to the 2~µm period of the wire pattern.

We additionally note that for spatially homogeneous systems, the momentum-resolved noise power spectral density is mathematically equivalent to the ensemble average of the absolute square of the Fourier transform of each magnetic noise image~\cite{wiener_generalized_1930, noauthor_materials_nodate}. Thus, instead of having to compute every pair-wise correlation, which has time complexity of $\mathcal{O}(Mn^2)$, where $M$ is the number of shots and $n$ is the number of pixels, we can simply ensemble average the absolute square of the two-dimensional Fourier transform of each frame, which has time complexity $\mathcal{O}(Mn\log{n})$~\cite{lattuada_hitchhikers_2025}. This provides a significant speed-up in data processing, and becomes especially crucial with increased field of view
size.

By the properties of the Fourier transform, the minimum $q$ resolvable is given by the imaging field of view, which in our current configuration is approximately 4~µm, limited by the available laser power for spin-dependent ionization. If relevant, lower wavevectors can readily be accessed using more laser power or green readout. On the higher end, the maximum $q$ in principle is given by the pixel spacing, but is practically limited by optical diffraction (approximately 250~nm in our system). Thus, in its implementation here, the wide-field scheme unambiguously identifies noise signatures with spatial scales from $q_{\text{min}}=2\pi/L=2\pi\times0.25$~µm$^{-1}$ to $q_{\text{max}}=\pi/\delta x=2\pi\times2$~µm$^{-1}$. Longer wavelength features contribute solely to the noise power at $q=0$.

\section{Tunable momenta beyond the diffraction limit}
Thus far, we have used wide-field imaging of magnetic field correlations at many spatial separations to access the spatial structure of magnetic noise at diffraction-limited length scales. Here, we explore a new possibility to access fluctuation momenta beyond the diffraction limit by measuring correlations between NV centers in a single confocal spot.

In NV center noise spectroscopy, the fluctuation momenta $q$ in the sensing target that couple to a single NV center are given by the geometry of the experiment, and are quantified by the momentum filter function. While the exact form of this function depends on the nature and dimensionality of the fluctuation sources~\cite{agarwal_magnetic_2017, dolgirev_characterizing_2022, machado_quantum_2023}, the general behavior is captured by $W_d(q)\sim{}q^ke^{-2qd}$, where $d$ is the distance of the NV center to the noise source, and $k$ is related to the distance scaling of the field amplitude arising from the specific nature of the source~\cite{noauthor_materials_nodate}. Thus, the sensor-sample distance can be used to tune the fluctuation momenta that the NV center sees, enabling construction of a momentum spectrum. Here, we show that for a dense ensemble of NV centers, we can also tune the momentum that the ensemble is sensitive to by changing the lateral size of the sensing volume.

Our method enables measurement of the momentum spectrum of magnetic noise at fixed sensor-sample distance by measuring correlations between NV centers in a sensing volume with tunable size. The phase covariance between multiple NV centers in a single spot changes the variance of the measured photon distribution and can be extracted explicitly using linear combinations of phase-cycled variance measurements~\cite{rovny_multi-qubit_2025, noauthor_materials_nodate}. However, reading out the ensemble imposes spatial summing that averages out correlations with spatial periodicity smaller than the lateral size of the ensemble. Quantitatively, for an ensemble of sensors with sensor-sample distance $d$ in a Gaussian spot with diameter $D$, the momentum filter function is
\begin{equation}
	W_d^{\text{ens}}(q, D) \sim q^k\, e^{-2qd}\, \exp\!\left(-\alpha\,q^2 D^2\right)
	\label{eq:ensemble_mom_filt}
\end{equation}
where $\alpha$ is a constant of order one~\cite{noauthor_materials_nodate}. Combining this with super-resolution techniques to shrink the sensing volume below the diffraction limit~\cite{chen_subdiffraction_2015, mosavian_super-resolution_2024}, we can continuously tune the momentum filter function in a range~[\cref{fig:momentumfilter}(d)], analogous to moving a single NV center in a retractable scanning tip.

The schematic of our experiment is shown in \cref{fig:momentumfilter}(a). We use a diamond with a high-density of NV centers approximately 10$\pm$3~nm from the surface, and leverage two different crystallographic orientations of NV centers to achieve independent microwave control over the two subpopulations; this enables the phase cycling necessary to extract the phase covariance from photon variance measurements. Using a confocal microscope, we first initialize the spin and charge states of all NV centers using a green laser pulse, then use a high energy red doughnut pulse to deplete NV centers in a ring, leaving a Gaussian sensing volume with diameter $D$. We then apply a phase-cycled sensing sequence that measures correlations between the NV centers~\cite{rovny_multi-qubit_2025, noauthor_materials_nodate}, and read out the spin state using spin-to-charge conversion. While we cannot measure $D$ directly, we estimate the size of the sensing volume after depletion by measuring the point spread function of individually resolvable NV centers with the same doughnut beam~\cite{noauthor_materials_nodate}.

The spin state readout noise $\sigma_R$ increases with decreasing sensing volume~[\cref{fig:momentumfilter}(b)]. With readout noise characterized, we now turn to measuring correlated noise at tunable length scales below the diffraction limit. In this experiment, we apply global noise with an RF coil and sense the covariance between NV centers as a function of doughnut energy and thus changing momentum filter function, down to a diameter of approximately 100~nm~[\cref{fig:momentumfilter}(c)]. With decreasing diameter, the momentum filter function amplitude increases, but total covariance decreases and error increases because of the increasing readout noise. The use of two independently controlled orientations can also be used to measure temporal correlators at short time scales in small sensing volumes with high-$q$ momentum filter functions~\cite{noauthor_materials_nodate}.

The achievable tunable momentum filter function range in this scheme is comparable to a single NV center in a retractable scanning tip~[\cref{fig:momentumfilter}(d)]. However, the two schemes have different sensitivity tradeoffs and dynamic range~[\cref{fig:momentumfilter}(e)]. The single NV center noise measurement is a variance measurement whose sensitivity scales with a single power of $\sigma_R$, while the ensemble scheme is a correlation measurement whose sensitivity scales as $\sigma_R^2$. This means that while the single center sensitivity at close distances is better than the ensemble scheme, the sensitivity falls off drastically with increasing distance. In the ensemble case, as the momentum filter function moves to lower $q$, and thus lower amplitude, the readout noise improves, while at high $q$, the decreased NV center numbers and higher readout noise competes with the increased momentum filter function amplitude to flatten out the sensitivity.

The complementary NV center ensemble wide-field and sub-diffraction schemes open the door to probe tunable momentum regimes. The wide-field scheme enables direct access to the power spectral density of magnetic noise with spatial features on the order of 0.25~µm and longer. The sub-diffraction scheme enables continuous access down to approximately 100~nm length scales, but, like a single NV center, is unable to probe spatial anisotropy.

\section{Concluding remarks}
By accessing the spatial structure of magnetic noise, this work bridges a long-standing gap in materials spectroscopy. While scattering techniques remain the gold standard for mapping atomic-scale momentum responses at high-energy~\cite{gelmukhanov_dynamics_2021, gardner_high-resolution_2020}, these measurements are bulk-averaged and do not have access to low-energy, long-wavelength fluctuations where emergent phenomena such as phase separation and electron hydrodynamics occur~\cite{rovny_nanoscale_2024}. Our platform combines momentum resolution at these energy scales with the ability to probe nanoscale volumes down to the atomic monolayer limit~\cite{ku_imaging_2020, zhou_spatiotemporal_2020}. This enables direct visualization of strongly correlated matter, offering a platform to study the breakdown of transport regimes~\cite{agarwal_magnetic_2017, kolkowitz_probing_2015}, map the real-space divergence of correlation lengths across second-order phase transitions~\cite{huang_revealing_2023, machado_quantum_2023}, and isolate the microscopic mechanisms of dissipation, time-reversal symmetry breaking and order parameter fluctuations in superconductors~\cite{zhang_detecting_2026, zhang_nanoscale_2026, de_nanoscale_2026, hosseinabadi_theory_2026, romare_probing_2026, orgad_signatures_2026}. Beyond its immediate application to understanding materials and condensed matter phenomena, this multi-scale correlation architecture also forms the basis for several modalities of next-generation, quantum-enhanced metrology. The parallelized measurement of two-point correlators naturally extends to the extraction of higher-order, non-Gaussian spatial correlation functions~\cite{manning_third-order_2013}. Further, integrating quantum projection noise-limited readout~\cite{maier_readout_2026, bechelli_spin_2026} or engineering quantum correlated states within the ensemble using dipolar coupling~\cite{rovny_multi-qubit_2025, wu_spin_2025, gao_signal_2025} could dramatically amplify sensitivity, pushing momentum-resolved noise spectroscopy deep into the sub-diffraction regime.

During the preparation of this manuscript we became aware of a related paper~\cite{gao_nanoscale_2026} which leverages interactions between NV centers to selectively amplify correlated noise at wavevectors below the diffraction limit.

\begin{figure*}[h!]
	\centering
	\includegraphics[width=183mm]{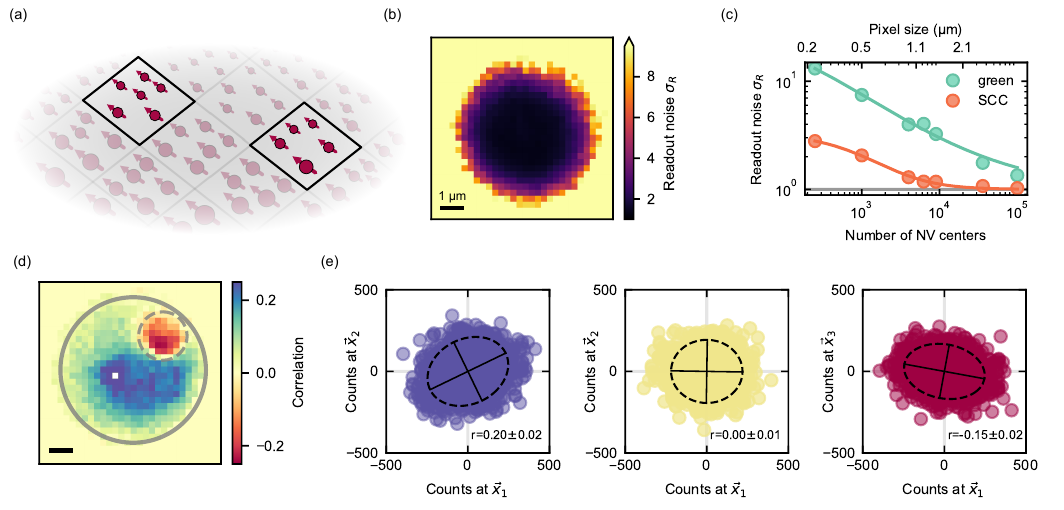}
	\caption{\textbf{High-fidelity, wide-field readout enables massively multiplexed correlation measurements.}
		(a) Schematic of camera readout of a high-density NV center ensemble.
		(b) Map of spin state readout noise $\sigma_R$, where each pixel is used as an individual sensor.
		(c) The readout noise $\sigma_R$ scales with the number of NV centers in the pixel. Spin-to-charge conversion (SCC) strongly outperforms green readout for low numbers of NV centers. The lines are fit to a power law lower bounded by 1.
		(d) Correlation map resulting from global driven Rabi oscillations. Sensors in the dashed circle are prepared in the opposite spin state to those in the solid circle only.
		(e) Photon count distributions (mean-subtracted) for different pixels sensing global noise using dynamical decoupling. For two pixels prepared in the same state ($\vec{x}_1$ and $\vec{x}_2$) the correlation coefficient is positive (blue); for pixels prepared in opposite states ($\vec{x}_1$ and $\vec{x}_3$) it is negative (red). Off resonance with the noise, no correlation is observed (yellow). Dashed lines are ellipse fits to the distributions; solid lines are their major and minor axes.}
	\label{fig:wfsccandcorrs}
\end{figure*}

\begin{figure*}[h!]
	\centering
	\includegraphics[width=183mm]{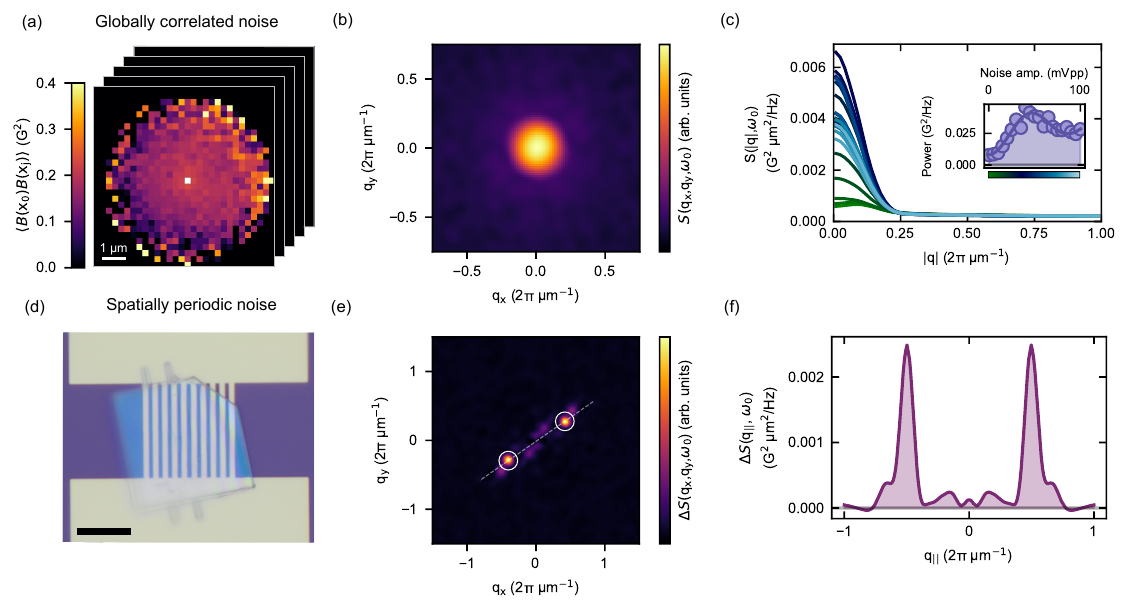}
	\caption{\textbf{Momentum-resolved magnetic noise spectroscopy reveals the spatial structure of magnetic noise.}
		(a) Two-point magnetic noise correlations at each pixel with respect to the white pixel across the imaging field of view, caused by globally applied noise from an RF coil.
		(b) The Fourier transform of multiplexed correlator measurements results in the momentum-resolved magnetic noise spectral density $S(\mathbf{q},\omega_0)$.
		(c) Radially averaged $S(|q|,\omega_0)$ profiles (color indicates applied noise amplitude). A Gaussian fit to each returns FWHM$_{|q|}\approx 2\pi\times0.2$~µm$^{-1}$, corresponding to the inverse of the imaging field of view (5~µm). Inset: total noise power from integrating each profile from $q=0$ to $q_{\text{max}}$; the line is a fit to the expected scaling of correlated noise amplitude.
		(d) Noisy current passed through an array of parallel wires modulates the magnetic field with spatial periodicity given by the wire spacing. A diamond microchip is placed on the device. Scale bar, 10~µm.
		(e) Momentum-resolved magnetic noise power spectral density $\Delta S(\mathbf{q},\omega_0)$ showing excess magnetic noise with applied current.
		(f) Line cut of $\Delta S(q_{||},\omega_0)$ reveals the spatial periodicity of
		the noise at $|q_{||}|\approx2\pi\times0.5$~µm$^{-1}$.}
	\label{fig:sqw}
\end{figure*}

\begin{figure*}[h!]
	\centering
	\includegraphics[width=183mm]{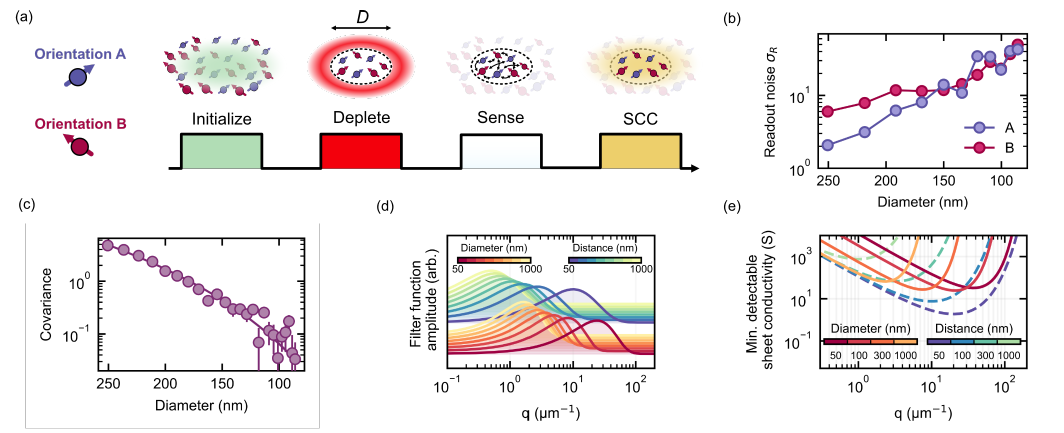}
	\caption{\textbf{Tunable momentum filter functions beyond the diffraction limit.}
		(a) Measuring correlations between NV center orientations A and B below the diffraction limit. Confocal lasers initialize the charge and spin states; a doughnut pulse then ionizes NV centers in a ring, leaving sensors within a diameter $D$. A sensing sequence measures correlations within the volume and is read out by spin-to-charge conversion.
		(b) Spin state readout noise $\sigma_R$ for each orientation scales strongly with decreasing sensing volume diameter.
		(c) Covariance between orientations A and B in a single spot, caused by applied noise at 2.5~MHz, versus sensing volume. Error bars, standard error of the variance for Gaussian distributions; line, power-law fit $D^{4.7\pm0.1}$.
		(d) Normalized momentum filter functions versus momentum $q$ for a two-dimensional sheet current, showing the overlap between the ensemble scheme with increasing diameter (red to orange) and the single NV center scheme with increasing sensor-sample distance (blue to green). Curves are each normalized to themselves and offset vertically.
		(e) Minimum detectable sheet conductivity in 10$^4$~s for the ensemble scheme (reds, solid) at a fixed NV center depth of 10~nm, and a single NV center at variable distances (blues, dashed).}
	\label{fig:momentumfilter}
\end{figure*}


\begin{acknowledgments}
We gratefully acknowledge helpful conversations with Artur Lozovoi, Shimon Kolkowitz, Jeff Thompson, Tian-Ming Fu, Zachary Krebs, Hiroto Takahashi, Dominick Corradino, and Divik Verma. We also thank Bart Machielse and Daniel Riedel (IonQ) for development of the diamond membranes. We additionally acknowledge Alex Healey for help with forming gas activation annealing of the bulk diamond crystal used in this work.

This work was supported by the Gordon and Betty Moore Foundation (GBMF12237, DOI 10.37807), the National Science Foundation (Grant No. NSF PHY-2514890), the Princeton Catalysis Initiative, and the Intelligence Community Postdoctoral Research Fellowship Program by the Oak Ridge Institute for Science and Education (ORISE) through an interagency agreement between the US Department of Energy and the Office of the Director of National Intelligence
(ODNI) (ZK \& JR).


The authors declare no competing financial interests.

\end{acknowledgments}

\pagebreak
\clearpage

\bibliography{references}

\clearpage
\onecolumngrid

\appendix

\begin{center}
\textbf{\large Supplemental material: Momentum-resolved quantum noise spectroscopy using ensembles of diamond quantum sensors}
\\
Z. Kazi et al.
\end{center}

\setcounter{equation}{0}
\setcounter{figure}{0}
\setcounter{table}{0}
\renewcommand{\thetable}{S\arabic{table}}
\renewcommand{\thefigure}{S\arabic{figure}}
\renewcommand{\theequation}{S\arabic{equation}}

\renewcommand{\thesection}{\arabic{section}}

\section{Materials and methods}

\subsection*{Diamond samples}

We use two types of diamond samples for the experiments. The first diamond sample, used for
all experiments except Fig.~3(d-f), was implanted with a nitrogen ion energy of 6~keV at
$1\times10^{13}$~cm$^{-2}$ and annealed in forming gas~\cite{healey_production_2026} then
oxygen terminated~\cite{sangtawesin_origins_2019}, resulting in an NV center ensemble
approximately 10~nm from the surface, and $T_{2,\text{XY8}}\approx20$~µs. The second
sample, used for the experiments shown in Fig.~3(d-f), is a diamond microchip from IonQ
Inc.~\cite{riedel_scalable_2026} that was implanted with a nitrogen energy of 85~keV at
$1\times10^{13}$~cm$^{-2}$, resulting in an NV center ensemble approximately 100~nm from the
surface, with $T_{2,\text{XY8}}\approx40$~µs.

\subsection*{Wide-field imaging setup}

\begin{figure}[h!]
	\centering
	\includegraphics[width=180mm]{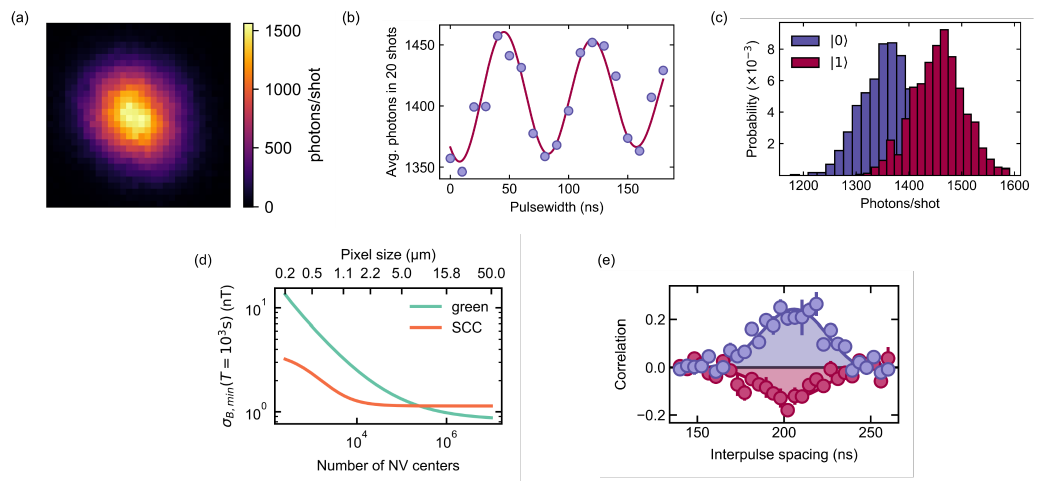}
	\caption{\textbf{Photon counts and readout noise using wide-field SCC.}
		(a) Wide-field orange laser spot showing counts per 1~ms readout.
		(b) Average Rabi oscillations measured at one pixel in 20 shots.
		(c) Histogram of counts at a single 250~nm pixel prepared in the $|0\rangle$
		and $|1\rangle$ state, with readout noise $\sigma_R\approx1.5$.
		(d) Minimum detectable correlated field amplitude in 1000~s as a function of
		number of NV centers. For large numbers, green readout results in better sensitivity
		because of the faster readout time.
		(e) Spectroscopy of a correlated applied noise signal. The pixel size in this
		experiment is 500~nm. The lines are fits to the expected correlation for noise
		sensing~\cite{rovny_nanoscale_2022}. The fitted root mean square correlated field
		amplitude is approximately 0.1~G, and the different amplitude between data sets is
		caused by the different values of readout noise for both sets of sensors. Error bars
		indicate the standard deviation across correlation values, computed by sampling a
		subset of the total number of experiment shots.}
	\label{fig:counts}
\end{figure}

The wide-field measurements shown in Figs.~2 and 3 were performed in a home-built setup that
combines wide-field green, orange, and red laser illumination, combined with a confocal
imaging path using a polarizing beam splitter (CCM1-PBS251). The sample is measured using a
100x, NA = 1.3 oil-immersion objective (Nikon N100X-PFO). The wide-field green laser is a
532~nm diode-pumped solid state laser (Coherent Verdi G5) modulated by an acousto-optic
modulator (AOM) (AA Optics MCQ110-A2-VIS). The orange laser used for NV center charge state
readout is a 594~nm, diode-pumped solid state laser (Coherent Obis) modulated by an AOM (AA
Optics MCQ110-A2-VIS). The green and orange lasers are fiber coupled and combined (Thorlabs
RGB26HF). The red laser used for NV center ionization is a 640~nm diode-pumped solid state
laser (Cobolt Rogue) modulated by an AOM (AA Optics MCQ110-A2-VIS). The green/orange and red
are combined by using a dichroic mirror (Thorlabs DMSP 605). The readout path is separated
from the excitation by a dichroic mirror (Semrock FF647-SDi01). The NV center fluorescence is
long-pass filtered and then imaged onto a qCMOS camera (Hamamatsu OrcaQuest). The confocal
green laser used for spin re-initialization is a 520~nm fiber-pigtailed diode laser (Cobolt
06 MLD). The green and orange laser illumination areas are approximately 6~µm across, and
laser powers are approximately 20~mW and 1~mW, respectively. The green spin and charge
initialization time is 100~µs and the orange charge state readout time is 1~ms, resulting
in about 1500 photons per shot in a wide-field spot (\cref{fig:counts}). The camera
readout time is approximately 400~µs. The red spot size is approximately 4~µm across,
and total power before the objective is approximately 850~mW. The ionization pulse time is
250~ns. The confocal re-initialization beam power is approximately 50~µW, and
re-initialization time is 45~µs. The wide-field noise sensing experiments (Fig.~2(e)) were
done using an XY8-1 sequence with applied RF at 2.5~MHz.

The wide-field measurements shown in Fig.~3(d-f) were done in a comparable wide-field
imaging path that used a dry 100X NA = 0.9 objective (Olympus MPLFLN100X). The green and
orange lasers were combined in free space using a dichroic mirror (Thorlabs DMSP567), then
combined with the red laser with another dichroic (Thorlabs DMSP605). The readout path is
separated from the excitation by another dichroic mirror (Thorlabs DMSP650), and the
fluorescence is filtered and imaged onto the same camera. The green and orange laser
illumination areas are approximately 30~µm across, and laser powers are approximately
40~mW and 7~mW, respectively. The green spin and charge initialization time is 3~ms and the
orange charge state readout time is 1~ms. The camera readout time is approximately
500~µs. The red spot size is approximately 10~µm across, and total power before the
objective is approximately 880~mW. The ionization pulse time is 250~ns. The wide-field noise
sensing experiments (Fig.~3(d-f)) were done using an XY8-3 sequence with applied RF at
2.5~MHz, and peak-to-peak current in the wire was approximately 1~mA.

\subsection*{Confocal setup}

The confocal experiments shown in Fig.~4 were done using the oil immersion branch. The green
laser is the same 532~nm diode-pumped solid state laser (Coherent Verdi G5), modulated by an
AOM (AA Optics MCQ110-A2-VIS). The orange laser used is a 594~nm HeNe laser (Newport)
modulated by an AOM (Isomet M1205-P80L-1). The red lasers used for charge state depletion and
ionization for readout are directly modulated diode lasers (Cobolt 06-MLD 638~nm). The
doughnut beam is linearly polarized (Thorlabs LPVISC100-MP2), shaped using a vortex wave
plate (Thorlabs WPV10L-633), and circularly polarized (Thorlabs WPQ10M-588). The green laser
power is approximately 80~µW, the orange power is approximately 10~µW, and both
ionization and doughnut depletion powers are approximately 15~mW. The correlated noise
sensing experiment was done with an XY8 sequence with applied RF at 2.5~MHz. The applied
magnetic field was aligned to enable independent addressing of two NV center crystallographic
orientations.

\subsection*{Microwave control}

Microwave pulses are generated using a signal generator (Rohde \& Schwarz SMATE200A),
amplified (Mini-Circuits ZHL-16W-43S+), and sent to a homemade microwave printed circuit
board~\cite{yuan_instructional_2024} for air-immersion measurements or a coverslip-stripline
for oil-immersion measurements. IQ phase modulation is achieved using arbitrary waveform
generators (AWGs) (Keysight 33622A) that modulate the different quadratures of the signal
generator. Phase-random noise is generated by another AWG (Keysight 33622A).

\section{Correlations measured between NV center ensembles}

Here we derive the measured Pearson correlation between NV center ensembles. The derivation
largely follows from Rovny et al.~\cite{rovny_nanoscale_2022}, but the use of ensembles
enables a small gain in sensitivity due to reduced projection noise.

\subsection*{Setup}

We measure two separate ensembles of NV centers by detecting their total brightness on two
pixels. Each NV center is initialized in the transverse plane with an initial $\pi/2$ pulse,
and acquires a random phase $\phi$ which is drawn from a zero-mean distribution. We make no
assumptions about the shape of this distribution. A final $\pi/2$ pulse out of phase with the
first converts the transverse-plane phase $\phi$ into a polar phase $\theta=\phi+\pi/2$.
Below, we write everything in terms of $\phi$. At readout, each NV center will be found in
state $m_{s}=0$ or $m_{s}=1$.

We assume the number of photons detected from each NV center are drawn from a Poisson
distribution. The number of NV centers found to be in $m_{s}=0$ is described by a Binomial
distribution. We define $k$ as the number of NV centers in state $m_s=0$, $M$ as the number
of NV centers in each pixel, $\phi$ as the phase acquired by each NV center, and $\alpha_0,
\alpha_1$ as the mean photon number from a single NV center found in state $m_{s}=0,1$.

The mean value of $k$ across iterations of measuring phase $\phi$ is~\cite{itano_quantum_1993}
\begin{equation}
\langle k\rangle_{\phi}=\sum_{k=0}^{M}k~p(k)=\int_{\phi}\left[\sum_{k=0}^{M}k~p(k|\phi)\right]p(\phi)d\phi
\end{equation}
where
\begin{equation}
p(k|\phi)=\frac{M!}{k!(M-k)!}p_{\uparrow}(\phi)^k(1-p_{\uparrow}(\phi))^{M-k}
\end{equation}
\begin{equation}
p_{\uparrow}(\phi)=\cos^{2}\left(\frac{\theta}{2}\right)=\cos^{2}\left(\frac{\phi+\pi/2}{2}\right)=\frac{1}{2}(1-\sin(\phi))
\end{equation}
and $p(\phi)$ is the distribution of $\phi$ which depends on the pulse sequence and signal
source (e.g. random-phase AC).

The mean value of a Binomial distribution is just $\langle k\rangle=Mp$ so the mean value in
the sum (the term in brackets above) evaluates to
\begin{equation}
\sum_{k=0}^{M}k~p(k|\phi)=Mp_{\uparrow}(\phi)
\end{equation}
This same kind of trick allows us to make similar substitutions any time we can say something
about the known statistics of given distributions (Binomial, Poisson).

\subsection*{Ideal readout}

With ideal readout, the Pearson correlation is:
\begin{equation}
r=\frac{\langle k_{1}k_{2}\rangle-\langle k_{1}\rangle\langle k_{2}\rangle}{\sigma_{k_{1}}\sigma_{k_{2}}}
\end{equation}

First, we get the numerator. The mean squared value and individual means are
\begin{align}
\langle k_{1}k_{2}\rangle_{\phi} &= \int_{\phi}\left[\sum_{k_{1}=0}^{M}k_{1}p(k_{1}|\phi_{1})\right]\left[\sum_{k_{2}=0}^{M}k_{2}p(k_{2}|\phi_{2})\right]p(\phi_{1},\phi_{2})d\phi \nonumber \\
&= M^{2}\langle p_{\uparrow}(\phi_{1})p_{\uparrow}(\phi_{2})\rangle=\frac{1}{4}M^{2}\left(1+\langle \sin(\phi_{1})\sin(\phi_{2})\rangle\right)
\end{align}
\begin{equation}
\langle k_{1}\rangle\langle k_{2}\rangle=M^{2}\langle p_{\uparrow}(\phi_{1})\rangle\langle p_{\uparrow}(\phi_{2})\rangle=\frac{1}{4}M^{2}
\end{equation}
so that the numerator is $\frac{1}{4}M^{2}\langle \sin(\phi_{1})\sin(\phi_{2})\rangle$.

The variance of the counts from one cluster of M spins is a little harder, since the Binomial
draws are now self-correlated (drawn from the self-same distribution). The variance of a
Binomial distribution is $\mathrm{Var}_{B}\equiv\langle k^{2}\rangle-\langle
k\rangle^{2}=Mp(1-p)$, where $\langle k\rangle=Mp$ so $\langle
k^{2}\rangle=Mp(1-p)+M^{2}p^{2}$. The problem is that the variance after using many values of
$\phi$ instead of just a single value for $p$ is that the center of the binomial distribution
is changing, so the variance of the net distribution is higher. Thus, we calculate variance
by doing the integral over $\phi$. First, note that for evenly-distributed $\phi$ (noise
sensing) we have $\langle k\rangle=\int M~p_{\uparrow}(\phi)p(\phi)d\phi=M/2$ (see the
equation for $p_{\uparrow}(\phi)$ above):
\begin{align}
\langle k^{2}\rangle-\langle k\rangle^{2} &= \int_{\phi}(Mp(1-p)+M^{2}p^{2})p(\phi)d\phi-\frac{1}{4}M^{2} \nonumber \\
&= \frac{1}{2}M-\frac{1}{4}M^{2}+(M^{2}-M)\int\frac{1}{4}(1-\sin(\phi))^{2}p(\phi)d\phi \nonumber \\
&= \frac{1}{4}\left[M^{2}\langle \sin^{2}(\phi)\rangle+M\langle \cos^{2}(\phi)\rangle\right]
\end{align}

For instance, if $\phi=0$, then we recover $\mathrm{Var}_B=M/4$ which is the expected
variance for a Binomial distribution with $p=1/2$.

Then the correlation is
\begin{equation}
r=\frac{\langle \sin(\phi_{1})\sin(\phi_{2})\rangle}{\left[\langle \sin^{2}(\phi_{1})\rangle+\frac{1}{M}\langle \cos^{2}(\phi_{1})\rangle\right]^{1/2}\left[\langle \sin^{2}(\phi_{2})\rangle+\frac{1}{M}\langle \cos^{2}(\phi_{2})\rangle\right]^{1/2}}
\end{equation}

For large M this is
\begin{equation}
r\approx\frac{\langle \sin(\phi_{1})\sin(\phi_{2})\rangle}{\sqrt{\langle \sin^{2}(\phi_{1})\rangle}\sqrt{\langle \sin^{2}(\phi_{2})\rangle}}=\frac{\langle \sin(\phi_{1})\sin(\phi_{2})\rangle}{\sin(\phi_{1})_{\text{rms}}\sin(\phi_{2})_{\text{rms}}}
\end{equation}

For perfectly correlated phases, the correlation is 1 for large M instead of $1/2$ for $M=1$.
An easier way to understand the bound $r\le1$:
\begin{equation}
r^{2}\approx\frac{\langle \sin(\phi_{1})\sin(\phi_{2})\rangle}{\langle \sin^{2}(\phi_{1})\rangle}\frac{\langle \sin(\phi_{1})\sin(\phi_{2})\rangle}{\langle \sin^{2}(\phi_{2})\rangle}
\end{equation}

\subsection*{Including shot noise}

We now seek
\begin{equation}
r=\frac{\langle N_{1}N_{2}\rangle-\langle N_{1}\rangle\langle N_{2}\rangle}{\sigma_{N_{1}}\sigma_{N_{2}}}
\end{equation}
where $N_i$ are the photon distributions measured at each pixel. We will calculate the
numerator and denominator separately.

\subsubsection*{The numerator}

The terms in the numerator are straightforward to calculate, since the Poisson draws are
independent:
\begin{align}
\langle N\rangle &= \sum_{k}\int_{N,\phi}N~p(N|k)p(k|\phi)p(\phi)d\phi~dN \nonumber \\
&= \sum_{k}\int_{\phi}(k\alpha_{0}+(M-k)\alpha_{1})p(k|\phi)p(\phi)d\phi \nonumber \\
&= \alpha_{0}\int_{\phi}Mp_{\uparrow}(\phi)p(\phi)d\phi+\alpha_{1}\int_{\phi}M(1-p_{\uparrow}(\phi))p(\phi)d\phi \nonumber \\
&= \frac{1}{2}M\alpha_{0}+M\alpha_{1}\left(1-\frac{1}{2}\right)=\frac{1}{2}M(\alpha_{0}+\alpha_{1})
\end{align}

\begin{align}
\langle N_{1}N_{2}\rangle &= \sum_{k_{1},k_{2}}\int_{N_{1},N_{2},\phi_{1},\phi_{2}}N_{1}N_{2}p(N_{1}|k_{1})p(N_{2}|k_{2})p(k_{1}|\phi_{1})p(k_{2}|\phi_{2})p(\phi_{1},\phi_{2})d\phi~dN \nonumber \\
&= \sum_{k_{1},k_{2}}\int_{\phi_{1},\phi_{2}}[k_{1}\alpha_{0}+(M-k_{1})\alpha_{1}][k_{2}\alpha_{0}+(M-k_{2})\alpha_{1}]p(k_{1}|\phi_{1})p(k_{2}|\phi_{2})p(\phi_{1},\phi_{2})d\phi \nonumber \\
&= \int_{\phi_{1},\phi_{2}}[M(\alpha_{0}-\alpha_{1})p_{\uparrow,1}(\phi)+M\alpha_{1}][M(\alpha_{0}-\alpha_{1})p_{\uparrow,2}(\phi)+M\alpha_{1}]p(\phi_{1},\phi_{2})d\phi \nonumber \\
&= M^{2}(\alpha_{0}-\alpha_{1})^{2}\langle p_{\uparrow,1}p_{\uparrow,2}\rangle+M^{2}(\alpha_{0}-\alpha_{1})\alpha_{1}+M^{2}\alpha_{1}^{2}
\end{align}

Then the numerator is
\begin{equation}
\langle N_{1}N_{2}\rangle-\langle N_{1}\rangle\langle N_{2}\rangle=\frac{1}{4}M^{2}(\alpha_{0}-\alpha_{1})^{2}\langle \sin(\phi_{1})\sin(\phi_{2})\rangle
\end{equation}

\subsubsection*{The denominator}

We want to calculate the variance of the photon counts from a single pixel,
$\sigma_{N}^{2}=\langle N^{2}\rangle-\langle N\rangle^{2}$, where the second term is easily
found from $\langle N\rangle=\frac{1}{2}M(\alpha_{0}+\alpha_{1})$.

We now calculate the first term, the mean squared photon count:
\begin{align}
\langle N^{2}\rangle &= \sum_{k}\int_{N,\phi}N^{2}p(N|k)p(k|\phi)p(\phi)d\phi~dN \nonumber \\
&= \sum_{k}\int_{\phi}[k\alpha_{0}+(M-k)\alpha_{1}][1+k\alpha_{0}+(M-k)\alpha_{1}]p(k|\phi)p(\phi)d\phi \nonumber \\
&= \sum_{k}\int_{\phi}[(\alpha_{0}-\alpha_{1})^{2}k^{2}+(\alpha_{0}-\alpha_{1})(1+2M\alpha_{1})k+M^{2}\alpha_{1}^{2}+M\alpha_{1}]p(k|\phi)p(\phi)d\phi \nonumber \\
&= \int_{\phi}\left[(\alpha_{0}-\alpha_{1})^{2}[(M^{2}-M)p_{\uparrow}^{2}+Mp_{\uparrow}]\right. \nonumber \\
&\qquad \left.+(\alpha_{0}-\alpha_{1})(1+2M\alpha_{1})Mp_{\uparrow}+M^{2}\alpha_{1}^{2}+M\alpha_{1}\right]p(\phi)d\phi \nonumber \\
&= [(\alpha_{0}-\alpha_{1})^{2}(M^{2}-M)]\langle p_{\uparrow}^{2}\rangle \nonumber \\
&\quad +[(\alpha_{0}-\alpha_{1})M(1+2M\alpha_{1})+(\alpha_{0}-\alpha_{1})^{2}M]\langle p_{\uparrow}\rangle+M\alpha_{1}+M^{2}\alpha_{1}^{2} \nonumber \\
&= \frac{1}{2}(\alpha_{0}+\alpha_{1})M+M^{2}\alpha_{0}\alpha_{1}+\frac{1}{4}(\alpha_{0}-\alpha_{1})^{2}\left[M\langle \cos^{2}(\phi)\rangle+M^{2}(\langle \sin^{2}(\phi)\rangle+1)\right]
\end{align}

The variance is
\begin{equation}
\sigma_{N}^{2}=\langle N^{2}\rangle-\langle N\rangle^{2}=\frac{1}{2}M(\alpha_{0}+\alpha_{1})+\frac{1}{4}(\alpha_{0}-\alpha_{1})^{2}\left[M\langle \cos^{2}(\phi)\rangle+M^{2}\langle \sin^{2}(\phi)\rangle\right]
\end{equation}

\subsection*{Correlation with shot noise}

Assuming both photon distributions have the same variance for simplicity, the correlation is
\begin{align}
r &= \frac{\frac{1}{4}M^{2}(\alpha_{0}-\alpha_{1})^{2}\langle \sin(\phi_{1})\sin(\phi_{2})\rangle}{\frac{1}{2}M(\alpha_{0}+\alpha_{1})+\frac{1}{4}(\alpha_{0}-\alpha_{1})^{2}\left[M\langle \cos^{2}(\phi)\rangle+M^{2}\langle \sin^{2}(\phi)\rangle\right]} \nonumber \\
&= \frac{\langle \sin(\phi_{1})\sin(\phi_{2})\rangle}{\frac{1}{M}\langle \cos^{2}(\phi)\rangle+\langle \sin^{2}(\phi)\rangle+\frac{2}{M}\frac{(\alpha_{0}+\alpha_{1})}{(\alpha_{0}-\alpha_{1})^{2}}}
\end{align}

So the correlation and effective readout noise are
\begin{equation}
r=\frac{1}{\sigma_{R,M}^{2}}\langle \sin(\phi_{1})\sin(\phi_{2})\rangle
\end{equation}
\begin{equation}
\sigma_{R,M}=\sqrt{\frac{1}{M}\langle \cos^{2}(\phi)\rangle+\langle \sin^{2}(\phi)\rangle+2\frac{(M\alpha_{0}+M\alpha_{1})}{(M\alpha_{0}-M\alpha_{1})^{2}}}
\end{equation}

Compare to the typical single-NV center expression~\cite{rovny_nanoscale_2022}:
\begin{equation}
\sigma_{R}=\sqrt{1+2\frac{(\alpha_{0}+\alpha_{1})}{(\alpha_{0}-\alpha_{1})^{2}}}
\end{equation}

The readout noise simplifies depending on the brightness of the distribution. For $M\gg1$:
\begin{equation}
\sigma_{R,M}^{2}\xrightarrow{M\gg1}\langle \sin^{2}(\phi)\rangle+2\frac{(M\alpha_{0}+M\alpha_{1})}{(M\alpha_{0}-M\alpha_{1})^{2}}
\end{equation}

For $M\gg(\alpha_{0}+\alpha_{1})/(\alpha_{0}-\alpha_{1})^{2}\approx1250$ (for typical readout
noise $\sigma_{R}=50$):
\begin{equation}
\sigma_{R,M}^{2}\rightarrow\langle \sin^{2}(\phi)\rangle
\end{equation}

\subsection*{Arbitrary photon count distributions}

Often, the photon count distribution for an NV center measurement is not Poissonian, for
instance because the ionization step in spin-to-charge conversion is not perfectly selective,
or because the counts are read out using a camera with a complicated response function.

Using $n_{i}$ to denote the number of photons detected from NV $i$ where $i=1...M$, we have
\begin{equation}
\langle N^{2}\rangle=\left\langle\left(\sum_{i}n_{i}\right)^{2}\right\rangle=M\langle n^{2}\rangle+(M^{2}-M)\langle n_{i}n_{j}\rangle
\end{equation}

Calculating the two terms on the right:
\begin{equation}
\langle n^{2}\rangle=\sum_{m}\int_{n,\phi}n^{2}p(n|m)p(m|\phi)p(\phi)d\phi~dn=\frac{1}{2}(\langle n_{0}^{2}\rangle+\langle n_{1}^{2}\rangle)=\frac{1}{2}(\alpha_{0}^{2}+\sigma_{0}^{2}+\alpha_{1}^{2}+\sigma_{1}^{2})
\end{equation}
\begin{equation}
\langle n_{i}n_{j}\rangle=\frac{1}{4}(1+\langle \sin^{2}(\phi)\rangle)(\alpha_{0}-\alpha_{1})^{2}+\alpha_{0}\alpha_{1}
\end{equation}
where we used the expression for the variance $\sigma^{2}=\langle n^{2}\rangle-{\langle
n\rangle}^{2}$ and the previously-derived value for $\langle n_{1}n_{2}\rangle$
(see~\cite{rovny_nanoscale_2022}).

The variance is then
\begin{equation}
\mathrm{Var}=\langle N^{2}\rangle-\langle N\rangle^{2}=\frac{1}{4}M(\alpha_{0}-\alpha_{1})^{2}\left[\langle \cos^{2}(\phi)\rangle+M\langle \sin^{2}(\phi)\rangle+2\frac{\sigma_{0}^{2}+\sigma_{1}^{2}}{(\alpha_{0}-\alpha_{1})^{2}}\right]
\end{equation}

and the correlation is
\begin{equation}
r=\frac{\langle \sin(\phi_{1})\sin(\phi_{2})\rangle}{\left(\sigma_{R,M}^{\mathrm{gen}}\right)^{2}}
\end{equation}
\begin{align}
\sigma_{R,M}^{\mathrm{gen}} &= \sqrt{\frac{1}{M}\langle \cos^{2}(\phi)\rangle+\langle \sin^{2}(\phi)\rangle+2\frac{M\sigma_{0}^{2}+M\sigma_{1}^{2}}{(M\alpha_{0}-M\alpha_{1})^{2}}} \nonumber \\
&\approx \sqrt{\langle \sin^{2}(\phi)\rangle+2\frac{M\sigma_{0}^{2}+M\sigma_{1}^{2}}{(M\alpha_{0}-M\alpha_{1})^{2}}}
\end{align}

\section{Correlation from parametric photon distributions}

Each panel of Fig.~\ref{fig:wfsccandcorrs}(e) contains a parametric plot of mean-subtracted photon counts $\delta N_i = N_i - \bar{N}_i$ recorded simultaneously at positions $\vec{x}_i$ over $M$ repetitions. We approximate this joint distribution as a Gaussian,
$P(\delta\vec{N}) \propto \exp[-\tfrac{1}{2}\,\delta\vec{N}^{\mathsf T}
\Sigma^{-1} \delta\vec{N}]$, with covariance 
\begin{equation}
  \Sigma =
  \begin{pmatrix} \sigma_1^2 & r\,\sigma_1\sigma_2 \\
                  r\,\sigma_1\sigma_2 & \sigma_2^2 \end{pmatrix},
\end{equation}
where $r$ is the Pearson correlation and $\sigma_i$ are the standard deviations of the photon counts at each location. The principal axes of the fitted ellipses to these distributions are the eigenvectors of $\Sigma$, with semi-axes $\propto \sqrt{\lambda_\pm}$ and
tilt $\theta$ given by
\begin{equation}
  \lambda_\pm = \frac{\sigma_1^2+\sigma_2^2}{2}
  \pm \sqrt{\Big(\frac{\sigma_1^2-\sigma_2^2}{2}\Big)^{2}
  + r^2\sigma_1^2\sigma_2^2},
  \qquad
  \tan 2\theta = \frac{2\,r\,\sigma_1\sigma_2}{\sigma_1^2-\sigma_2^2}.
  \label{eq:ellipse}
\end{equation}
Inverting these equations nets the Pearson correlation. For the data shown in Fig.~2E, the blue distribution gives $\theta$~=~0.45~rad and the red distribution gives $\theta~=$~-0.18~rad.

Note that for pixel pairs of equal total variance, $\sigma_1=\sigma_2=\sigma$,
Eq.~\eqref{eq:ellipse} reduces to $\lambda_\pm = \sigma^2(1\pm r)$ with
principal axes along $\pm \pi/4$: the sign of $r$ fixes the
direction of the tilt, while its magnitude is set by the eccentricity,
\begin{equation}
  |r| = \frac{a^2-b^2}{a^2+b^2},
\end{equation}
where $a$ and $b$ are the semi-major and semi-minor axes.

\section{Constructing the momentum-resolved magnetic noise spectral density}

The momentum-resolved magnetic noise spectral density is most generally defined as
\begin{equation}
    S(\mathbf{q},\omega) = \langle B(\mathbf{q},\omega)B(-\mathbf{q}, -\omega)\rangle.
\end{equation}
Assuming a two-dimensional, translationally invariant system, we can expand the right hand
side as
\begin{equation}
    \langle B(\mathbf{q},\omega)B(-\mathbf{q}, -\omega)\rangle = \int d^2\boldsymbol{\ell}\ \int dt\
    e^{-i\mathbf{q}\cdot \boldsymbol{\ell}}e^{i\omega t}\langle B(\mathbf{x}_i,t)B(\mathbf{x}_j, 0)\rangle
\end{equation}
where $\boldsymbol{\ell}=\mathbf{x}_i-\mathbf{x}_j$. NV center schemes can simplify the
temporal fluctuations using a sharp filter function, either with the use of dynamical
decoupling sequences or relaxometry. In the case of sensing a phase-random AC field at
frequency $\omega_0$ with dynamical decoupling~\cite{degen_quantum_2017}, we have

\begin{equation}
    \langle B(\mathbf{q},\omega_0)B(-\mathbf{q}, -\omega_0)\rangle =
    \frac{N\pi}{\omega_0}\int d^2\boldsymbol{\ell}\
    e^{-i\mathbf{q}\cdot \boldsymbol{\ell}}\langle B(\mathbf{x}_i)B(\mathbf{x}_j)\rangle,
    \label{eq:FTofcorrfunc}
\end{equation}

\noindent\ which is \cref{eq:sqw} in the main text. Notably, this result is equivalent to the
ensemble average of the absolute square of the two-dimensional Fourier transform of
individual magnetic field maps~\cite{lattuada_hitchhikers_2025}:
\begin{equation}
S(\mathbf{q}, \omega_0) =
    \frac{N\pi}{\omega_0}\int d\mathbf{x}_i\int d\mathbf{x}_j\
    e^{-i\mathbf{q}\cdot \mathbf{x}_i}e^{i\mathbf{q}\cdot \mathbf{x}_j}\langle B(\mathbf{x}_i)B(\mathbf{x}_j)\rangle
\end{equation}
\begin{align}
    =\frac{N\pi}{\omega_0}\langle \int d\mathbf{x}_i\ e^{-i\mathbf{q}\cdot \mathbf{x}_i}
     B(\mathbf{x}_i)\int d\mathbf{x}_j\ e^{i\mathbf{q}\cdot \mathbf{x}_j}B(\mathbf{x}_j)\rangle
\end{align}
\begin{align}
    =\frac{N\pi}{\omega_0}\langle|\int d\mathbf{x}_j\
    e^{-i\mathbf{q}\cdot \mathbf{x}_j}B(\mathbf{x}_j)|^2\rangle
\end{align}
\begin{align}
S(\mathbf{q}, \omega_0) = \frac{N\pi}{\omega_0}\langle|\mathcal{F}\{B(\mathbf{x})\}|^2\rangle.
\end{align}
In practice, this object is computationally easier to calculate, providing a significant
speed-up in data processing when dealing with large number of repetitions or large image
datasets.

In our experiment, we measure Pearson correlations between NV center ensembles, and we can
use a previous result to go from independent measurements of Pearson correlation, readout
noise and decoherence to get~\cite{cheng_massively_2025}
\begin{equation}
    S(\mathbf{q},\omega_0) = \frac{\delta x^2\ \omega_0}{\gamma_e^2 N\pi \bar{W^2}}\sum_{i\neq j}
    e^{-i\mathbf{q}\cdot \mathbf{\Delta x}}\sinh^{-1}\!\left(r_{ij}\frac{\sigma_{Ri}\sigma_{Rj}}{e^{-\chi_i(t)-\chi_j(t)}}\right)
\end{equation}
where $\delta x$ is the physical pixel size, $\gamma_e$ is the electron gyromagnetic ratio,
$N$ is the number of dynamical decoupling pulses, $\bar{W^2}$ is the peak amplitude of the
filter function~\cite{degen_quantum_2017}, $r_{ij}$ is the Pearson correlation between
location $\mathbf{x}_i$ and $\mathbf{x}_j$, $\sigma_{Ri}$ is the readout noise at pixel $i$,
and $\chi_i(t)$ is the decoherence function at pixel $i$.

\section{Momentum filter functions}

Our scheme uses shallow NV centers for momentum-resolved noise spectroscopy, by measuring
correlations between NV centers in a sensing volume and by using the Wiener-Khinchin theorem
in wide-field. Here, we state the momentum filter functions for these sensing schemes, which
modify the single NV center filter function.

\subsection*{Single NV center momentum filter functions}

The momentum filter function for sensing with a single NV center depends on the nature and
source of fluctuations, and has been derived elsewhere for several
cases~\cite{agarwal_magnetic_2017, kolkowitz_probing_2015, machado_quantum_2023}. It relates
the noise measured at an NV center to the fluctuations in a source material. In a simplified
form, we can write
\begin{equation}
S^{\text{single}}_B(\omega) = \int_0^\infty \frac{dq}{2\pi}\, W^{\text{single}}_d(q)\,
\mathcal{S}(q, \omega),
\label{eq:single_filt_def}
\end{equation}
where $\mathcal{S}(q, \omega)$ is the dynamical structure factor of the material. In what
follows, we use a simplified general form
\begin{equation}
    W^{\text{single}}_{d}(q) = W_0\; q^k\;e^{-2qd},
    \label{eq:general_filt}
\end{equation}
where the exponent $k$ is set by the character and dimensionality of the source, with specific examples
given in \cref{tab:table_singleNV_filtfunc}, and the unitful prefactor $W_0$ collects the material and geometric constants.

\begin{table}[htbp]
\begin{center}
\begin{tabular}{ll}
\\
\hline
Source & $W_{d}(q)$ \\
\hline
2D sheet dipole~\cite{dolgirev_characterizing_2022, machado_quantum_2023}    & $q^{3}e^{-2qd}$ \\
3D bulk dipole      & $q^{2}e^{-2qd}$  \\
2D sheet current~\cite{agarwal_magnetic_2017, dolgirev_characterizing_2022}   & $q^{1}e^{-2qd}$  \\
3D bulk current~\cite{kolkowitz_probing_2015}    & $q^{0}e^{-2qd}$ \\
\hline
\end{tabular}
\end{center}
\caption{\textbf{Momentum filter functions for varying source types and geometries.} The
	three-dimensional entries result from integrating the two-dimensional contribution over
	many uncorrelated layers.}
\label{tab:table_singleNV_filtfunc}
\end{table}

\subsection*{Single-volume ensemble momentum filter function}

Then, the magnetic noise measured by the ensemble
\begin{equation}
    S_{B}^{\text{ens}}(\omega) = \int_{-\infty}^{\infty} dt\, e^{i\omega t}\,
    \langle \bar{B}(t)\bar{B}(0)\rangle
\end{equation}
is modified by the NV density profile, giving
\begin{equation}
    S_{B}^{\text{ens}}(\omega) = \int_0^\infty \frac{dq}{2\pi}\;W_d^{\text{single}}(q)\,
    |\tilde{P}(q)|^2\,\mathcal{S}(q,\omega),
\end{equation}
which gives the momentum filter function for sensing in a single spot:
\begin{equation}
    W_d^{\text{ens}}(q) = W_d^{\text{single}}(q)\, |\tilde{P}(q)|^2.
\end{equation}

For a Gaussian profile with FWHM $D=2\sigma\sqrt{2\ln2}$, we have
\begin{equation}
    P(\boldsymbol{\rho}) = \frac{1}{{2\pi\sigma^2}}\exp\!\left(-\frac{|\boldsymbol{\rho}|^2}{2\sigma^2}\right),\; \tilde{P}(\mathbf{q}) = \exp\!\left(-\frac{|\mathbf{q}|^2\sigma^2}{2}\right) = \exp\!\left(-\frac{q^2 D^2}{16\ln 2}\right),
\end{equation}
resulting in
\begin{equation}
    W_d^{\text{ens}}(q, D) =  W_d^{\text{single}}(q) \exp\!\left(-\frac{q^2 D^2}{8\ln 2}\right),
    \label{mom_filter_ens}
\end{equation}
which is \cref{eq:ensemble_mom_filt} in the main text. In the limit $D\rightarrow0$, we recover the single NV
center result. For $D\gg d$, the Gaussian function cuts off the filter function at $q\sim1/D$
before the exponential term can suppress it. Effectively, this means the ensemble is more
sensitive to long-wavelength fluctuations and less sensitive to short wavelength fluctuations
than a single NV center at the same depth. However, the ability to tune $D$ in situ optically
allows for the ability to tune the momentum filter function peak without changing
sensor-sample distance.

\begin{figure}[h!]
    \centering
    \includegraphics[width=110mm]{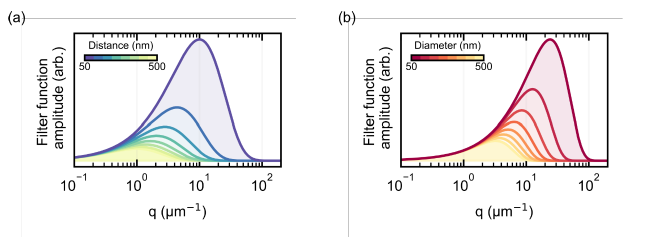}
    \caption{\textbf{Comparison of momentum filter functions.}
    	(a) Single NV center momentum filter function for sensing 2D currents as a
    	function of distance, normalized to the lowest distance curve.
    	(b) Ensemble NV center momentum filter function for sensing 2D currents as a
    	function of ensemble diameter, normalized to the lowest diameter curve. The NV-sample
    	distance is assumed to be 10~nm.}
    \label{fig:momentum_filt_not_norm}
\end{figure}

In our scheme, we use two NV center crystallographic orientations to enable independent
control for measuring fast time dynamics, and phase cycling to remove correlated background
fluctuations. In this scheme, the momentum filter function sampled is modified by a numerical
factor $\mathcal{G}^{AB} = \frac{1}{2}\hat{n}^A\cdot\hat{n}^B + \frac{1}{2}n^A_z n^B_z$ where
$n^A$ and $n^B$ are the unit vectors along the two NV center orientations:
\begin{equation}
    W_d^{AB}(q, D) = \mathcal{G}^{AB}\, W_d^{\text{single}}(q)\, \exp\!\left(-\frac{q^2 D^2}{8\ln 2}\right),
\end{equation}
where $\mathcal{G}^{AB} =1/3$ for the case of the NV orientations shown in the main text. The
tunable filter functions for sensing 2D sheet currents with single centers and NV center
ensembles are shown in \cref{fig:momentum_filt_not_norm}.

\subsection*{Momentum filter function for sensing between NV center ensembles}

The wide-field scheme is able to fully resolve the momentum spectrum of magnetic noise, which
is directly related to $\mathcal{S}(\mathbf{q},\omega)$. Each pixel is labeled by its position
$\mathbf{x}_i$, and the magnetic field measured using dynamically decoupled sensing at
$\omega_0=\pi/\tau$ at each pixel is $B_{\omega_0}(\mathbf{x}_i)$. Assuming translational
symmetry, the two-point correlation function for two positions separated by
$\boldsymbol{\ell}=\mathbf{x}_i-\mathbf{x}_j$ is $C_{\omega_0}(\boldsymbol{\ell})$. The ability
to sample many separations $\boldsymbol{\ell}$ enables a Fourier transform of the correlation
function,
\begin{equation}
S_{B}^{\text{wide-field}}(\mathbf{q},\omega_0) = \int d^2\ell\; e^{-i\mathbf{q}\cdot
\boldsymbol{\ell}}\, C_{\omega_0}(\boldsymbol{\ell}).
\label{eq:wf_def}
\end{equation}
We relate this to the single NV center filter function by noting that the noise sensed by a
single NV sensor is the zero-separation limit of the multi-point case:
\begin{equation}
S_B^{\text{single}}(\omega_0) = C_{\omega_0}(\mathbf{0}) = \int \frac{d^2q}{(2\pi)^2}\,
S^{\text{wide-field}}(\mathbf{q},\omega_0) = \int_0^\infty \frac{dq}{2\pi}\; q\,
S^{\text{wide-field}}(q,\omega_0),
\end{equation}
where the last step assumes the filter kernels in
\cref{tab:table_singleNV_filtfunc} depend only on $|q|$. Comparing with
\cref{eq:single_filt_def} identifies the wide-field filter function,
\begin{equation}
W_d^{\text{wide-field}}(q) = \frac{1}{q}\;W_d^{\text{single}}(q).
\label{eq:wf_filt}
\end{equation}
The wide-field scheme samples the same exponential standoff factor as a single NV center but
one power of $q$ lower. Unlike \cref{eq:single_filt_def}, \cref{eq:wf_filt} is a
pointwise relation, so it holds for anisotropic $\mathcal{S}(\mathbf{q},\omega_0)$ as well.

The final step includes the optical effects from wide-field readout: because we are reading
out the ensemble with diffraction-limited optics onto a camera pixel, the effective filter
function includes the optical point spread function and a top-hat spatial convolution from
the camera pixel shape. Writing $\mathcal{A}(\boldsymbol{\rho})$ for the resulting normalized
per-pixel sensing profile, the measured spectral density is
\begin{equation}
S^{\text{wide-field}}_{\text{meas}}(\mathbf{q},\omega_0) = |\tilde{\mathcal{A}}(\mathbf{q})|^2\,
W_d^{\text{wide-field}}(q)\, \mathcal{S}(\mathbf{q},\omega_0),
\end{equation}
where $\tilde{\mathcal{A}}(\mathbf{q})$ is the product of the incoherent optical transfer
function and the transform of the pixel area, proportional to $|P(q)|$ with width given by
the diffraction limit. In the momentum range accessible in the wide-field scheme, $qd \ll 1$
for shallow NV centers, so $e^{-2qd}\approx1$, reducing the inversion of the measured
$S_B^{\text{wide-field}}(\mathbf{q},\omega_0)$ to obtain $\mathcal{S}(\mathbf{q},\omega_0)$ to
a simple power law scaling with a prefactor given by the optical effects, rather than an
integral relation as in the single center or single volume case.

\section{Extracting signals from multiple orientations using phase cycling}

Different NV center crystallographic orientations experience different Zeeman splittings,
meaning they can be independently controlled with microwaves. This was used in Rovny et
al.~\cite{rovny_nanoscale_2022} to measure temporal correlations between two individual NV
centers at short (instrument-limited) delay times. With NV center ensembles, both
orientations are present at each sensing location, requiring four measurements to extract the
relevant correlations~\cite{rovny_multi-qubit_2025}.

\subsection*{Spatiotemporal correlations}

\begin{figure}[h!]
    \centering
    \includegraphics[width=180mm]{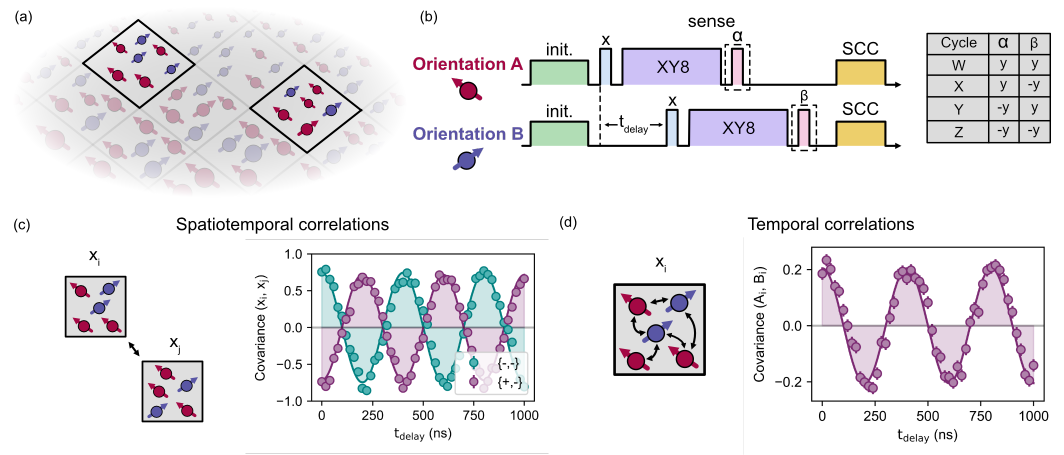}
    \caption{\textbf{Phase cycling to measure temporal correlations.}
    	(a) Schematic of ensemble showing two NV crystallographic orientations.
    	(b) Pulse sequence for measuring temporal correlations between NV center
    	ensembles. Both orientations are initialized, then a dynamical decoupling sequence is
    	applied to sense noise at a specific frequency. The start of the sensing sequence for
    	orientation B is offset by a delay time. After phase accumulation, different phase
    	pulses are used to map the population to different spin states, and the entire frame is
    	read out using spin-to-charge conversion. The phase cycles to isolate the covariance
    	between orientations are shown at the right.
    	(c) Spatiotemporal correlations measured between location $\mathbf{x}_i$ and
    	$\mathbf{x}_j$ as a function of $t_{\text{delay}}$ for phase-random noise at 2.5~MHz.
    	When the opposite NV center transition is addressed for orientation A, the covariance
    	changes sign as expected.
    	(d) Temporal correlations measured at a single location $\mathbf{x}_i$.}
    \label{fig:tempcorr}
\end{figure}

We denote the photon distribution measured at a location $\mathbf{x}_i$ as $I_i=A_i+B_i$
where $A_i$ and $B_i$ are the photon counts from each NV center orientation. Then, the
quantity we are interested in for spatiotemporal correlations is $\text{Cov}(A_i,B_j)$.
However, we can only measure
\begin{equation}
    \text{Cov}(I(\mathbf{x}_i),I(\mathbf{x}_j))=\text{Cov}(A_i + B_i, A_j+B_j).
\end{equation}
In order to access the relevant quantity, we follow Rovny et
al.~\cite{rovny_multi-qubit_2025} and perform four phase-cycled measurements as shown in
\cref{fig:tempcorr}(b). These four measurements are
\begin{equation}
    S_W=\text{Cov}(A_i,A_j)+ \text{Cov}(A_i,B_j)+\text{Cov}(B_i, A_j)+\text{Cov}(B_i, B_j),
\end{equation}

\begin{equation}
    S_X=\text{Cov}(A_i,A_j)+\text{Cov}(A_i,B'_j)+\text{Cov}(B'_i, A_j)+\text{Cov}(B'_i, B'_j),
\end{equation}

\begin{equation}
    S_Y=\text{Cov}(A'_i,A'_j)+\text{Cov}(A'_i,B_j)+\text{Cov}(B_i, A'_j)+\text{Cov}(B_i, B_j),
\end{equation}

\begin{equation}
    S_Z=\text{Cov}(A'_i,A'_j)+\text{Cov}(A'_i,B'_j)+\text{Cov}(B'_i, A'_j)+\text{Cov}(B'_i, B'_j),
\end{equation}
which flip the sign of the relevant covariance but leave the background terms untouched.
Then, we can linearly combine these measurements to obtain
\begin{equation}
    S_W+S_Z-S_X-S_Y=4[\text{Cov}(A_i,B_j)+\text{Cov}(B_i, A_j)]
\end{equation}
and
\begin{equation}
    S_W+S_Z+S_X+S_Y=4[\text{Cov}(A_i,A_j)+\text{Cov}(B_i, B_j)],
\end{equation}
and define
\begin{equation}
    \text{Covariance}(\mathbf{x}_i,\mathbf{x}_j)=\frac{S_W+S_Z-S_X-S_Y}{S_W+S_Z+S_X+S_Y}.
\end{equation}
We measure this quantity as a function of $t_{\text{delay}}$ for phase-random AC noise in
\cref{fig:tempcorr}(c).

\subsection*{Correlations within a single spot}

Similarly, for two NV center orientations at a single spot $x_i$, the variance of the photon
distribution is given by the sum:
\begin{equation}
    \text{Var}(x_i)=\text{Var}(A_i + B_i).
\end{equation}
Expanding, we have
\begin{equation}
    \text{Var}(x_i)=\text{Var}(A_i) + \text{Var}(B_i) + 2\ \text{Cov}(A_i,B_i).
\end{equation}

In the same way as above, we can perform four measurements and take linear combinations to
get
\begin{equation}
    S_W+S_Z-S_X-S_Y=8\ \text{Cov}(A_i,B_i)
\end{equation}
and
\begin{equation}
    S_W+S_Z+S_X+S_Y=4\ [\text{Var}(A_i)+\text{Var}(B_i)]
\end{equation}
and we define the quantity plotted in Fig.~4(c):
\begin{equation}
    C(A_i,B_i)=\frac{1}{8}(S_W+S_Z-S_X-S_Y).
\end{equation}

The error on this quantity is given by the error of a variance, assuming a Gaussian
distribution:
\begin{equation}
    \delta C(A_i,B_i)=\frac{1}{8}\sqrt{\frac{2}{M-1}(S_W^2 +S_X^2+S_Y^2+S_Z^2)},
\end{equation}
where $M$ is the number of shots per phase cycle.

\section{Sensitivity for measuring current fluctuations at varying $q$}

The sensitivity to a given dynamical structure factor inherits the $q$ dependence of the
momentum filter function. This means the sensitivity differs qualitatively between the source
types of \cref{tab:table_singleNV_filtfunc}. Here we work through the two-dimensional
sheet current case, $n=1$, relevant to Johnson noise in metals and to quasiparticle and
vortex dynamics in two-dimensional superconductors~\cite{kolkowitz_probing_2015, agarwal_magnetic_2017, dolgirev_characterizing_2022, zhang_nanoscale_2026}. The source is a
fluctuating sheet current density $\vec{\mathbf{J}}(\mathbf{r},t)$ in a layer at $z=0$, with
structure factor
\begin{equation}
  \mathcal{S}_{J}(\mathbf q,\omega)
  = \int\!\mathrm{d}t\,e^{i\omega t}\!\int\!\mathrm{d}^2r\;e^{-i\mathbf q\cdot\mathbf r}
    \bigl\langle \vec J(\mathbf r,t)\,\vec J(0,0)\bigr\rangle.
  \label{eq:SJJdef}
\end{equation}
For an NV center a distance $d$ above the sheet, we have the filter function
\begin{equation}
  W^{\text{single}}_{d}(q) = \frac{\mu_0^{2}}{4}\,q\,e^{-2qd}.
  \label{eq:W2Dcurrent}
\end{equation}
Following the literature, we transform the two-dimensional current structure factor to
in-plane conductivity $\mathrm{Re}\,[\sigma(q,\omega)]$. In the classical limit $k_B T_s \gg
\hbar\omega_0$, the fluctuation--dissipation theorem gives $\mathcal{S}_{J}(q,\omega) = 2k_B
T_s\,\mathrm{Re}\,[\sigma(q,\omega)]$, with $T_s$ the sample
temperature~\cite{kolkowitz_probing_2015}. We then integrate \cref{eq:single_filt_def}
over $q$, assuming the filter function has support in a band $\Delta q = q$ around $q$, which
gives
\begin{equation}
  S_{B}(\omega_0) \simeq \frac{q}{2\pi} W^{\text{single}}_{d}(q)\,\mathcal{S}_{J}(q,\omega_0)
  = \frac{\mu_0^{2} k_B T_s}{4\pi}\,q^{2}e^{-2qd}\,\mathrm{Re}\,[\sigma(q,\omega_0)].
  \label{eq:SB2D}
\end{equation}

To sense this noise we apply a dynamical decoupling sequence of $N$ pulses with interpulse
spacing $\tau$, which acts as a band-pass filter of width ${\sim}1/(N\tau)$ centered on
$\omega_{0}=\pi/\tau$. The variance of the accumulated phase is
\begin{equation}
  \bigl\langle \delta\phi^{2}\bigr\rangle
  = \eta\,\gamma_e^{2}\,N\tau\,S_{B}(\omega_{0}),
  \label{eq:phase}
\end{equation}
where $\eta = 4/\pi^{2}$ for a resonant $N$-pulse XY8 sequence. We note that depolarization
can also be used at this step to sense the magnetic noise~\cite{rovny_nanoscale_2024}.

At $\mathrm{SNR}=1$ the smallest resolvable phase variance is $\sigma_{R}e^{\chi}/\sqrt{M}$
for a single NV center variance measurement, and the smallest resolvable phase covariance is
$\sigma_{R,A}\sigma_{R,B}e^{\chi_A+\chi_B}/\sqrt{M} \approx \sigma_{R}^{2}e^{2\chi}/\sqrt{M}$
for the ensemble covariance measurement, where $\chi(t)=(t/T_2)^p$ is the decoherence
function evaluated at $t=N\tau$ and $M=T/(N\tau + t_{I}+t_{R})$ is the number of repetitions
in total measurement time $T$ with initialization and readout times $t_{I}$ and $t_{R}$.

Using this, we invert \cref{eq:SB2D} to give the minimum detectable conductivity at
momentum $q$:
\begin{align}
  \mathrm{Re}\,\sigma^{\text{single}}_{\min}(q)
  &= \frac{8\pi}{\eta\,\mu_0^{2}\,k_B T_s\,\gamma_e^{2}\,N\tau}\frac{1}{q^{2}e^{-2qd}}\;
     \frac{\sigma_{R}\,e^{\chi}}{\sqrt{M}}
  \label{eq:sigsingle}\\[4pt]
  \mathrm{Re}\,\sigma^{\text{ens}}_{\min}(q)
  &= \frac{4\pi}{\eta\,\mu_0^{2} k_B T_s \gamma_e^{2}\,N\tau}\, \frac{1}{\mathcal{G}^{AB}q^{2}e^{-2qd}
     |\tilde P(q)|^{2}}\;
     \frac{\sigma_{R}(D)^{2}\,e^{2\chi}}{\sqrt{M}}.
  \label{eq:sigens}
\end{align}
These are plotted in \cref{fig:momentumfilter}(e), using the following: $T_s = 300$~K, $N=8$, $\tau=1$~µs,
$\sigma_R^{\text{single}}=30$, $T_{2,\text{XY8}}^{\text{single}}=10$~µs (assuming a
shallow NV center in a scanning tip), $T_{2,\text{XY8}}^{\text{ensemble}}=20$~µs
(measured from the sample used in this work), $\mathcal{G}^{AB}=1/3$, $t_I=10$~µs,
$t_R^{\text{single}}=300$~ns, $t_R^{\text{ensemble}}=1$~ms, $T=10^4$~s.

\section{Diameter estimate from single NV center point spread function}

In our scheme, we use a depletion beam to shrink the sensing volume. We estimate the size of
the ensemble after depletion by calibrating to single NV center measurements with the same
doughnut beam~\cite{chen_subdiffraction_2015}. Here, the doughnut pulse time is 15~µs,
and the charge state readout is done at 10~µW for 2~ms.

\begin{figure}[h!]
    \centering
    \includegraphics[width=110mm]{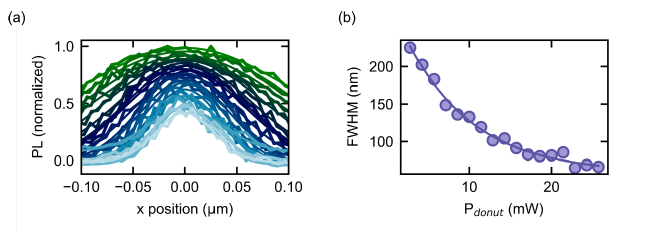}
    \caption{\textbf{Single NV center charge state depletion.}
    	(a) Line scans over a single center with increasing doughnut power (green to
    	light blue).
    	(b) Full width at half maximum as a function of doughnut power. The
    	exponential fit gives us a mapping between doughnut energy and sensing volume.}
    \label{fig:singleNVCSD}
\end{figure}

\end{document}